\documentclass[onecolumn,showpacs]{revtex4}

\usepackage{graphicx}
\usepackage{dcolumn}
\usepackage{bm}

\input epsf

\begin{document}

\title{\Large Correspondence between Ricci and other dark energies }

\author{\bf  Surajit
Chattopadhyay$^1$\footnote{surajit$_{_{-}}2008$@yahoo.co.in} and
Ujjal Debnath$^2$\footnote{ujjaldebnath@yahoo.com ,
ujjal@iucaa.ernet.in}}

\affiliation{$^1$Department of Computer Application, Pailan
College of Management and Technology, Bengal Pailan Park,
Kolkata-700 104, India.\\
$^2$Department of Mathematics, Bengal Engineering and Science
University, Shibpur, Howrah-711 103, India. }

\date{\today}

\begin{abstract}
Purpose of the present paper is to view the correspondence between
Ricci and other dark energies. We have considered the Ricci dark
energy in presence of dark matter in non-interacting situation.
Subsequently, we have derived the pressure and energy density for
Ricci dark energy. The equation of state parameter has been
generated from these pressure and energy density. Next, we have
considered the correspondence between Ricci and other dark energy
models, namely tachyonic field, DBI-essence and new agegraphic
dark energy without any interaction and investigated possible
cosmological consequences.
\\
\end{abstract}


\maketitle

\section{\normalsize\bf{Introduction}}

An exotic form of negative pressure matter called dark energy is
used to explain the acceleration of the universe inferred from the
the observations of distant type Ia supernovae, cosmic microwave
background radiation (CMBR), and Sloan Digital Sky Survey (SDSS)
[1]. The simplest candidate of dark energy is the vacuum energy
density or cosmological constant $\Lambda$, whose energy density
remains constant with time $\rho_{\Lambda} = \Lambda/8\pi G$ and
whose equation of motion is also fixed,
$w=p_{\Lambda}/\rho_{\Lambda}=-1$  ($p_{\Lambda}$ is the pressure)
during the evolution of the universe. The resulting cosmological
model, $\Lambda CDM$, consists of a mixture of vacuum energy and
cold dark matter. Another possibility is QCDM cosmologies based on
a mixture of cold dark matter and quintessence $(-1<w\leq0)$ [2].
To alleviate the 'fine tuning' and 'cosmological coincidence'
problems [3] associated with the $\Lambda CDM$, various dark
energy candidates have been proposed such as quintessence
mentioned, k-essence, tachyons, phantoms, ghost condensates and
quintom, etc. An important advancement in the studies of black
hole theory and string theory is the suggestion of the so-called
holographic principle, which may provide some clues for solving
these problems. Gao et al [4] has discussed how the holographic
dark energy model deals with the two fundamental problems
mentioned above. This holographic dark energy model and its
interacting versions are successful in fitting the current
observations [5]. Inspired by the holographic dark energy
models,Gao et al [4] proposed another possibility, where the
density is proportional to the Ricci scalar curvature $R$ and
named this dark energy as Ricci dark energy (RDE). The Ricci
scalar of FRW universe is given by
$R=-6(\dot{H}+2H^{2}+\frac{k}{a^{2}})$, where dot denotes a
derivative with respect to time $t$ and $k$
 is the spatial curvature.
\\

  The purpose of the present
work is to investigate the correspondence between Ricci dark
energy and other dark energy candidates namely tachyonic field
[6], DBI-essence [7] and new agegraphic dark energy [8]. Although
the RDE is a recently introduced dark energy model, a literature
survey shows that some significant works have been already done in
RDE, which can be regarded as a kind of holographic dark energy
with the square root of the inverse Ricci scalar as its infrared
cutoff [9]. The RDE is given by its density [4]

\begin{equation}
\rho_{R}=3 c^{2}\left(\dot{H}+2H^{2}+\frac{k}{a^{2}}\right)
\end{equation}

where $c$ is a dimensionless parameter which will determine the
evolution behavior of RDE. When $c^{2}<1/2$, the RDE will exhibit
a quintomlike behavior; i.e., its equation of state will evolve
across the cosmological-constant boundary $w=-1$ [4]. In the work
by Zhang under reference [4], $c^{2}$ was taken as 0.3, 0.4, 0.5,
and 0.6 and it was observed that for $c^{2}<1/2$ the equation of
state parameter $w<-1$, i.e. behaves like quintom and for
$c^{2}>1/2$ it has been shown that $w<-1$, i.e. behaves like a
quintessence. In flat FRW universe, where $k=0$, the RDE will be
given by

\begin{equation}
\rho_{R}=3 c^{2}(\dot{H}+2H^{2})
\end{equation}

For a flat FRW universe the Hubble parameter is given by
$H=\dot{a}/a$, where $a$, a function of the cosmic time $t$, is
the scale factor. Some recent advances in RDE are the references
in [10]. In our work, we would consider the RDE with dark matter
and would investigate its association with the other dark
energies.
\\

\section{\normalsize\bf{Ricci dark energy model with dark matter}}

In this section, we include the dark matter without interaction
with RDE. The conservation equations for RDE and dark matter are
written as

\begin{equation}
\dot{\rho}_{R}+3H
(\rho_{R}+p_{R})=0;~~~~~~\dot{\rho}_{m}+3H(\rho_{m}+p_{m})=0
\end{equation}

Solving the conservation equation for RDE we get

\begin{equation}
p_{R}=-c^{2}\left(\frac{\ddot{H}}{H}+7\dot{H}+6H^{2}\right)
\end{equation}

and for dark matter we introduce a constant equation of state
parameter $w_{m}$ so that

\begin{equation}
p_{m}=w_{m}\rho_{m}
\end{equation}

As we are considering a mixture of the RDE and dark matter we
write

\begin{equation}
\begin{array}{c}
  \rho_{1}=\rho_{R}+\rho_{m} \\
  p_{1}=p_{R}+p_{m} \\
\end{array}
\end{equation}

For the mixture of RDE and dark matter without interaction, the
conservation equation would take the form

\begin{equation}
\dot{\rho}_{1}+3H (1+w_{1})\rho_{1}=0
\end{equation}

where, $w_{1}=\frac{p_{R}+p_{m}}{\rho_{R}+\rho_{m}}$.

Using the conservation equation for dark matter and the equation
(5) we get

\begin{equation}
\rho_{m}=\rho_{om} a^{-(1+w_{m})}
\end{equation}

where the subscript $0$ denotes the value of $\rho_{m}$ at present
time (zero redshift). Using the expression for $w_{1}$ in the
conservation equation (7) we can rewrite the Friedman equation as
follows

\begin{equation}
H^{2}=\frac{1}{3(2+(w_{m}-4)c^{2})}\left[3a^{\frac{2}{c^{2}}-4}c_{1}((w_{m}-4)c^{2}+2)+2
a^{-w_{m}}c^{2}\rho_{om}\right]
\end{equation}

\begin{equation}
\begin{array}{c}
  \rho_{1}=\frac{1}{3(2+(w_{m}-4)c^{2})}\left[18a^{\frac{2}{c^{2}}-4}c_{1}c^{2}((w_{m}-4)c^{2}+2)
+12c^{4}\rho_{om}a^{-w_{m}}+~~~~~~~~~~~~~~~~~~~\right. \\
  \left.\frac{1}{c^{2}}\left(3\left(6a^{\frac{2}{c^{2}}-4}c_{1}c^{2}
(1-2c^{2})^{2}(2+(w_{m}-4)\right )
+c^{6}a^{-w_{m}}w_{m}^{2}\rho_{om}\right)
\sqrt{a^{\frac{2}{c^{2}}-4}c_{1}+\frac{2a^{-w_{m}}c^{2}\rho_{om}}{3(2+(w_{m}-4)c^{2})}}\right]
\end{array}
\end{equation}

\begin{equation}
p_{1}=\left(1-\frac{2}{c^{2}}\right )c_{1}a^{-4+\frac{2}{c^{2}}}-
\frac{c^{4}(w_{m}-4)(w_{m}-3)}{3(2+(w_{m}-4)c^{2})}\rho_{om}a^{-w_{m}}
\end{equation}
\\
\begin{figure}
\includegraphics[height=2.0in]{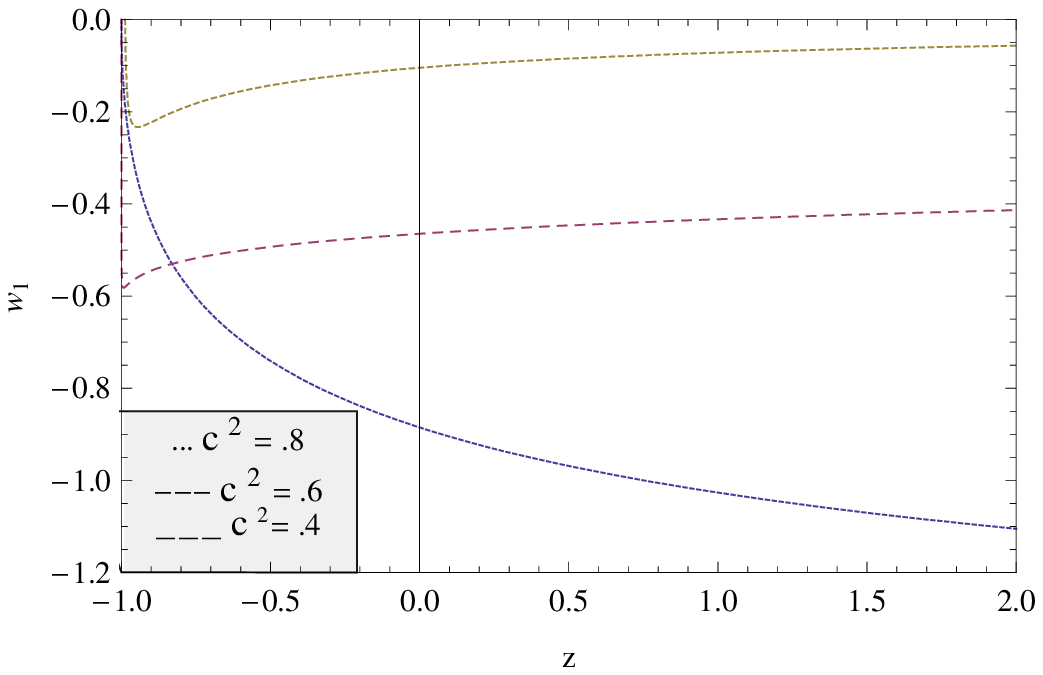}~\\
\vspace{1mm} ~~~~~~~~~~~~Fig.1\vspace{6mm} shows the variation of
the the equation of state parameter  $w_{1}$ against $z$ in the
case of Ricci dark energy model with dark matter.

\vspace{6mm}

\end{figure}
In figure 1, we have presented the evolution of the equation of
state parameter $w_{1}=p_{1}/\rho_{1}$ corresponding to the RDE
with dark matter. Since evolution of the equation of state
parameter for RDE is determined by the $c^{2}$, we have chosen it
below as well as above $1/2$. Our choices for $c^{2}$ are 0.4,
0.6, and 0.8. It is found in the figure 1 that when $c^{2}$ is 0.4
the equation of state parameter is exhibiting a quintomlike
evolution behavior, i.e. i.e., the equation of state of
holographic Ricci dark energy will evolve across the
cosmological-constant boundary $w_{1}=-1$ Actually, it evolves
from the region with $w_{1}<-1$ to that with $w_{1}> -1$ and this
transition implies a quintom model [11]. However, for $c^{2}$
greater than $1/2$ the equation of state always evolves above $-1$
and for $c^{2}=0.8$ the equation of state is tending to $0$ at
higher redshift and in implies that this RDE model is behaving
like dust matter during most of the epoch of matter domination.
\\
\section{\normalsize\bf{Correspondence between Ricci dark energy model
and tachyonic field }}

The present section aims to investigate the conditions under which
there is a correspondence between RDE model and the tachyonic
field, in the flat FRW Universe. That is, to determine an
appropriate potential for tachyonic field which makes the two dark
energies to coincide with each other. Let us first consider the
energy density $\rho$ and the pressure $p$ for the tachyonic field
[12]

\begin{equation}
\rho_{T}=\frac{V(\phi)}{\sqrt{1-{\dot{\phi}}^{2}}}
\end{equation}
and
\begin{equation}
p_{T}=-V(\phi) \sqrt{1-{\dot{\phi}}^{2}}
\end{equation}

for which the equation of state reads as

\begin{equation}
w_{T}=-(1-{\dot{\phi}}^{2})
\end{equation}

Equation $w_{T}$ with the RDE equation of state $w_{1}$ we
reconstruct the scalar field and potentials for the tachyonic
field as

\begin{equation}
   \begin{array}{c}
    \dot{\phi}^{2}=\frac{3a^{2-\frac{2}{c^{2}}}(c^{2}(w_{m}-4)+2)}{6c_{1}+c^{2}(3c_{1}w_{m}-12c_{1}+2\rho_{om}a^{4-\frac{2}{c^{2}}-w_{m}})}
[1+(3(c_{1}(c^{2}-2)(c^{2}(w_{m}-4)+2)-
\frac{1}{3}a^{4}c^{8}(w_{m}-4)(w_{m}-3)\rho_{om}))\\\\
    \times\left(18a^{\frac{2}{c^{2}}+w_{m}}c^{4}c_{1}(c^{2}(w_{m}-4)+2)+12a^{4}c^{6}\rho_{om}
3(6a^{\frac{2}{c^{2}}+w_{m}}c^{2
}c_{1}(c^{2}(w_{m}-4)+2)+a^{4}c^{6}w_{m}^{2}\rho_{om})\times\right.\\\\
     \left.\sqrt{c_{1}a^{-4+\frac{2}{c^{2}}}+
\frac{2c^{2}\rho_{om}a^{-w_{m}}}{3(c^{2}(w_{m}-4)+2
)}}\right)^{-1}]
   \end{array}
\end{equation}

and

\begin{equation}
\begin{array}{c}
 V=\frac{3c_{1}(c^{2}(w_{m}-4)+2)a^{-4+\frac{2}{c^{2}}}}{c^{2}(w_{m}-4)+2}\left[1-3a^{2+w_{m}}(c^{2}(w_{m}-4)+2)+
 \right.\\\\

\left(  \frac{9a^{6+2w_{m}}c^{4}(c^{2}(w_{m}-4)+2 )^{2}
 \left(\frac{(c^{2}-2)c_{1}a^{-4+\frac{2}{c^{2}}} }{c^{2}}+
 \frac{a^{-w_{m}}c^{4}(w_{m}-4)(w_{m}-3)\rho_{om} }
 {3(c^{2}(w_{m}-4)+2 ) } \right)}
{18a^{w_{m}+\frac{2}{c^{2}}}c^{4}c_{1}(c^{2}(w_{m}-4)+2 )+12a^{4}c^{6}\rho_{om}+
3(6a^{w_{}+\frac{2}{c^{2}}}c^{2}c_{1}(c^{2}(w_{m}-4)+2 )+a^{4}c^{6}w_{m}^{2}\rho_{om} )\times     }\right.\\\\

  \left.\left.\times \sqrt{ c_{1}a^{-4+\frac{2}{c^{2}}}+
\frac{2c^{2}\rho_{om}a^{-w_{m}}}{3(c^{2}(w_{m}-4)+2
)}}\right)\right]^{\frac{1}{2}}
\end{array}
\end{equation}

\begin{figure}
\includegraphics[height=2.0in]{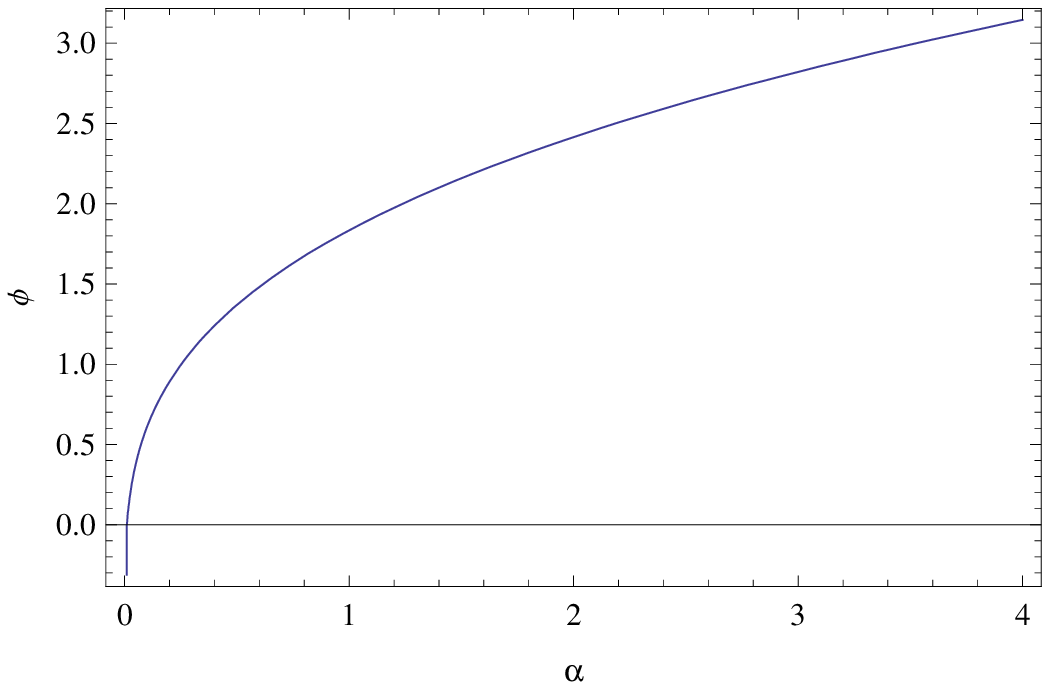}~~~
\includegraphics[height=2.0in]{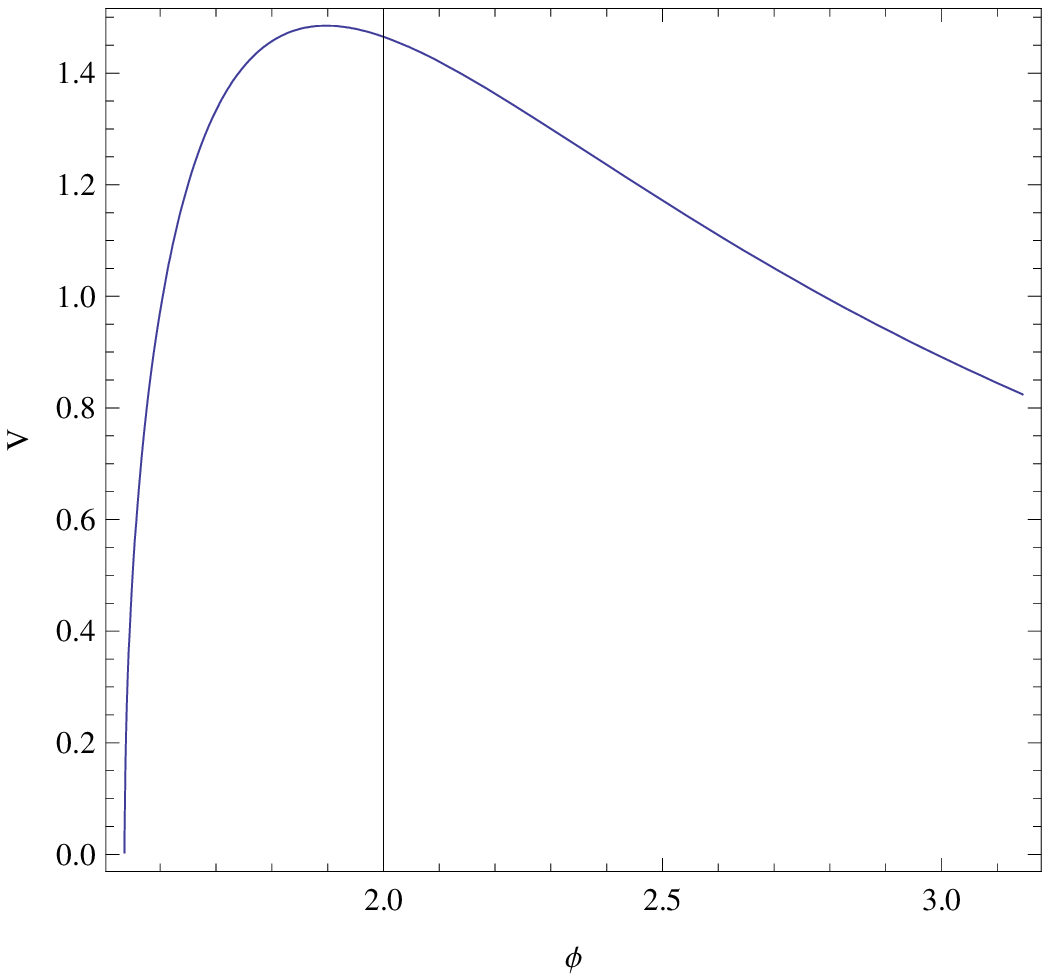}~\\
\vspace{1mm} ~~~~~~~~~~~~Fig.2~~~~~~~~~~~~~~~~~~~~~~~~~~~~~~~~~~~~~~~~~~~~~~~~~~~~~~~~Fig.3\\
\vspace{6mm} Figs. 2 and 3 show the variation of $\phi$ against
$a$  and $V$ against $\phi$ when a correspondence between RDE and
tachyonic field is considered. Here, $c^{2}=0.6$,
$\rho_{om}=0.01$, and $w_{m}=0.01$.

\vspace{6mm} \vspace{6mm}

\end{figure}

\begin{figure}
\includegraphics[height=2.0in]{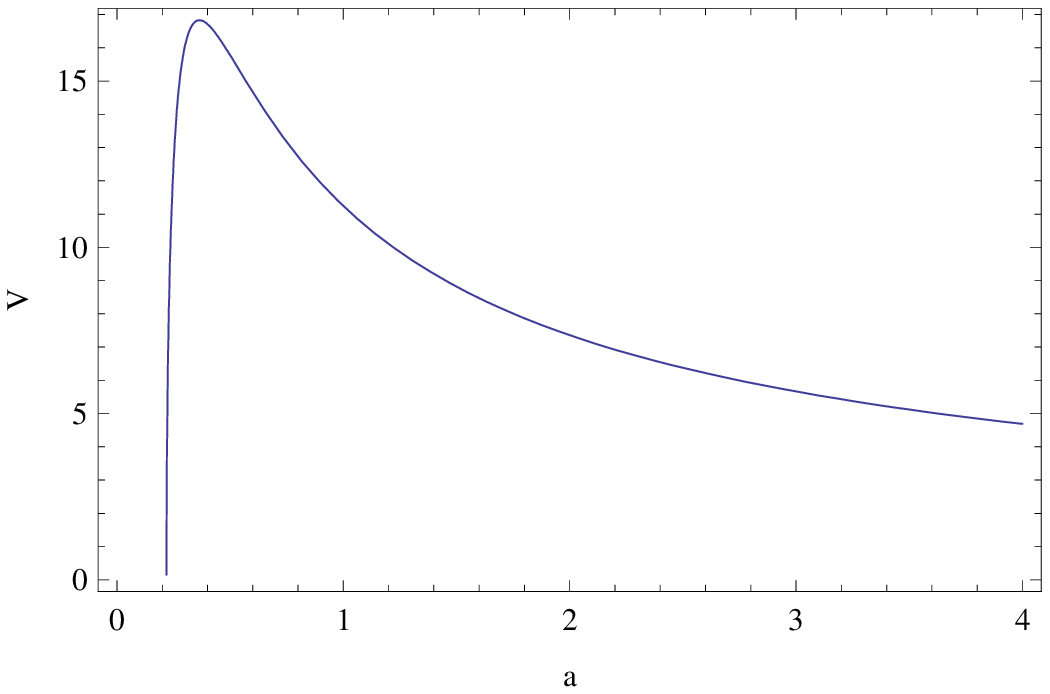}~\\
\vspace{1mm} ~~~~~~~~~~~~Fig.4\vspace{6mm} shows the variation of
the the potential  $V$ against $a$ when a correspondence between
RDE and tachyonic field is considered.

\vspace{6mm}

\end{figure}

When the field $\phi$ is plotted against the scale factor $a$, it
is observed that the field is increasing with increase in $a$. The
plot is presented in figure 2. When the potential $V$ is plotted
against the scalar field $\phi$ it is found that the potential is
decreasing with increase in the field $\phi$ as seen in figure 3.
In figure 4, we have seen that the potential of the tachyonic
field is decreasing with increase in the scale factor $a$. This
indicates that as the universe evolves, the potential of the
tachyonic field decreases when it is considered as Ricci dark
energy.
\\

\section{\normalsize\bf{Correspondence between Ricci dark energy model
and DBI essence }}

The density and pressure for DBI-essence are read as [13]

\begin{equation}
\rho=(\gamma-1)T(\phi)+V(\phi)
\end{equation}

\begin{equation}
p=\left(\frac{\gamma-1}{\gamma}\right)T(\phi)-V(\phi)
\end{equation}

where, $T(\phi)$ is the tension, $V(\phi)$ is the potential,
$\phi$ denotes the scalar field for DBI-essence and the quantity
$\gamma$ is reminiscent of the usual Lorentz factor given by

\begin{equation}
\gamma=\frac{1}{\sqrt{1-\frac{\dot{\phi}^{2}}{T(\phi)}}}
\end{equation}

\begin{figure}
\includegraphics[height=2.0in]{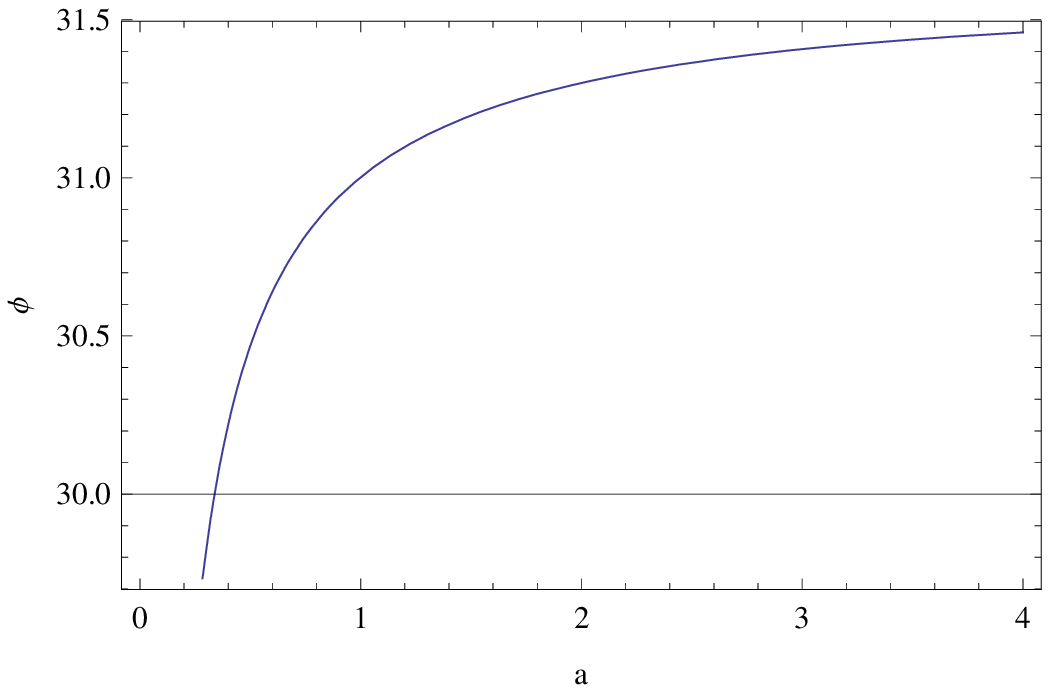}~~~

\includegraphics[height=2.0in]{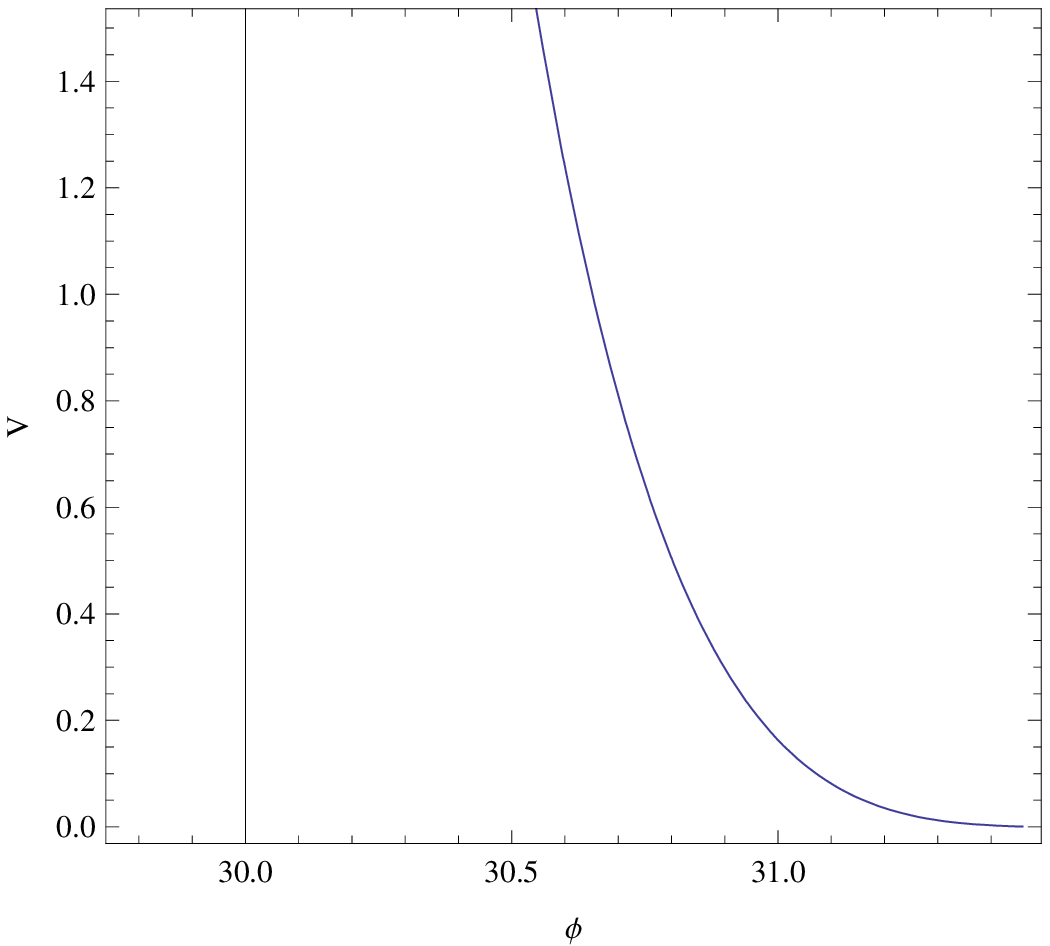}~\\
\vspace{1mm} ~~~~~~~~~~~~Fig.5~~~~~~~~~~~~~~~~~~~~~~~~~~~~~~~~~~~~~~~~~~~~~~~~~~~~~~~~Fig.6\\
\vspace{6mm} Figs. 5 and 6 show the variation of $\phi$ and $V$
against $a$ when DBI essence is considered as a Ricci dark energy.

\vspace{6mm}

\end{figure}

\begin{figure}
\includegraphics[height=2.0in]{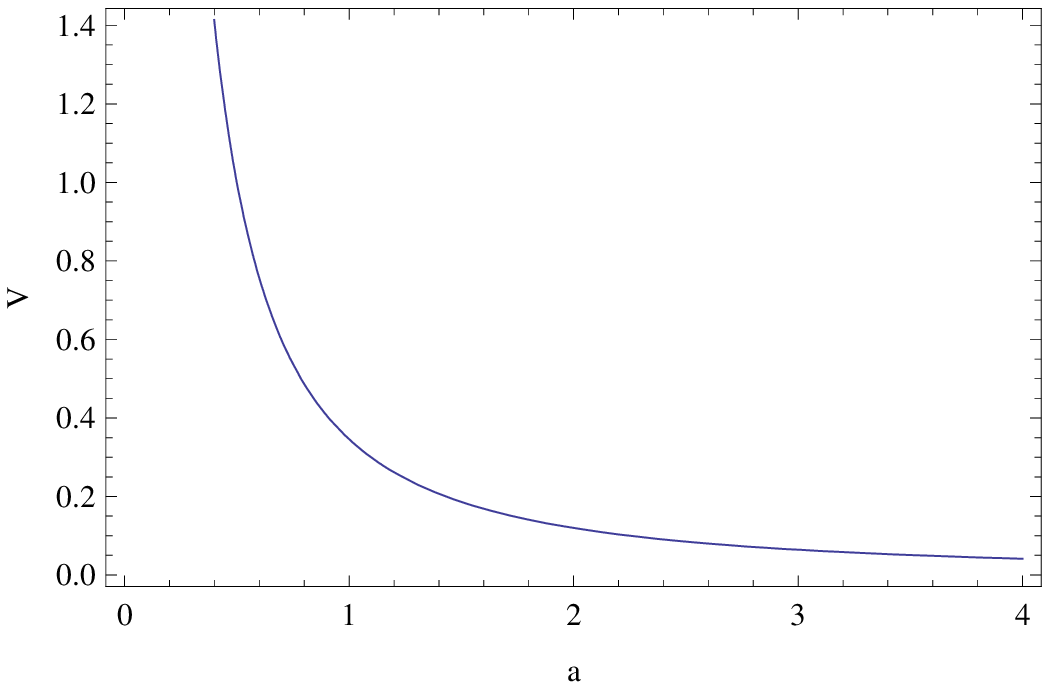}~\\
\vspace{1mm} ~~~~~~~~~~~~Fig.7\vspace{6mm} shows the variation of
the the potential  $V$ against $a$ when DBI essence is considered
as a Ricci dark energy.

\vspace{6mm}

\end{figure}

We assume that $T=n\dot{\phi}^{2}$, and we find that

\begin{equation}
\dot{\phi}^{2}=2\left(\frac{n-1}{n}\right)\left(\frac{(2c^{2}-1)c_{1}a^{-4+\frac{2}{c^{2}}}}{c^{2}}+\frac{c^{2}w_{m}\rho_{om}a^{-w_{m}}}{3(c^{2}(w_{m}-4)+2)}\right)
\end{equation}

\begin{equation}
V=2\frac{(\sqrt{n(n-1)}-w_{m})(\sqrt{n}-\sqrt{n-1})}{\sqrt{n-1}(w_{1}+1)}\left(\frac{n-1}{n}\right)\left(\frac{(2c^{2}-1)c_{1}a^{-4+\frac{2}{c^{2}}}}{c^{2}}+\frac{c^{2}w_{m}\rho_{om}a^{-w_{m}}}{3(c^{2}(w_{m}-4)+2)}\right)
\end{equation}

Where, $w_{1}$ is given by equation (13). It is observed in figure
5 that the scalar field of DBI essence is increasing with scale
factor $a$ when DBI essence in considered as Ricci dark energy. In
figure 6 it is seen that the potential of DBI essence in such case
is decreasing with increase in the scalar field $\phi$. The
decrease of the DBI essence potential with evolution of the
universe is apparent in figure 7.
\\

\section{\normalsize\bf{Correspondence between Ricci dark energy model
and new agegraphic dark energy }}

Quantum mechanics together with general relativity leads to the
  Karolyhazy relation and a corresponding energy density of quantum fluctuations of
space-time. Based on this energy density, Cai [14] proposed a dark
energy model, the so-called agegraphic dark energy model, in which
the age of the universe is introduced as the length measure. The
corresponding energy density is given by

\begin{equation}
\rho_{q}=\frac{3n^{2}m_{p}^{2}}{T^{2}}
\end{equation}

where,

\begin{equation}
T=\int\frac{da}{Ha}
\end{equation}

 The agegraphic dark energy was constrained by using some old high
redshift objects and type Ia supernovae. A new agegraphic dark
energy model was proposed in [15], where the time scale is chosen
as the conformal time $\eta$ instead of the age of the universe.
For this new agegraphic dark energy, the energy density $\rho_{A}$
is given as

\begin{equation}
\rho_{A}=\frac{3n^{2}m_{p}^{2}}{\eta^{2}}
\end{equation}

where

\begin{equation}
\eta=\int\frac{dt}{a}
\end{equation}

Thus, $\dot{\eta}=1/a$. The corresponding equation of state is

\begin{equation}
w=-1+\frac{2}{3n}\frac{\sqrt{\Omega_{A}}}{a}
\end{equation}

where,

\begin{equation}
\Omega_{A}=\frac{n^{2}}{H^{2}\eta^{2}}
\end{equation}

Taking the form of the Hubble parameter $H$ as in equation (9) and
$w$ as $w_{1}=p_{1}/\rho_{1}$ we get

\begin{equation}
\eta=\frac{3aH}{2}(1+rw_{1})
\end{equation}

\begin{figure}
\includegraphics[height=2.0in]{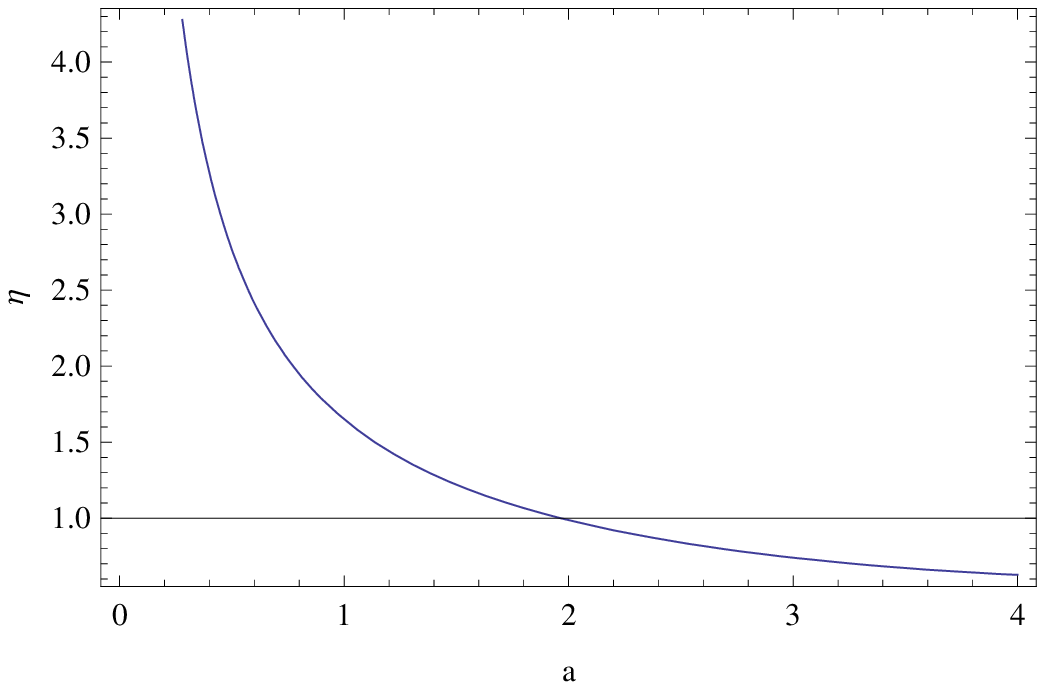}~~~
\includegraphics[height=2.0in]{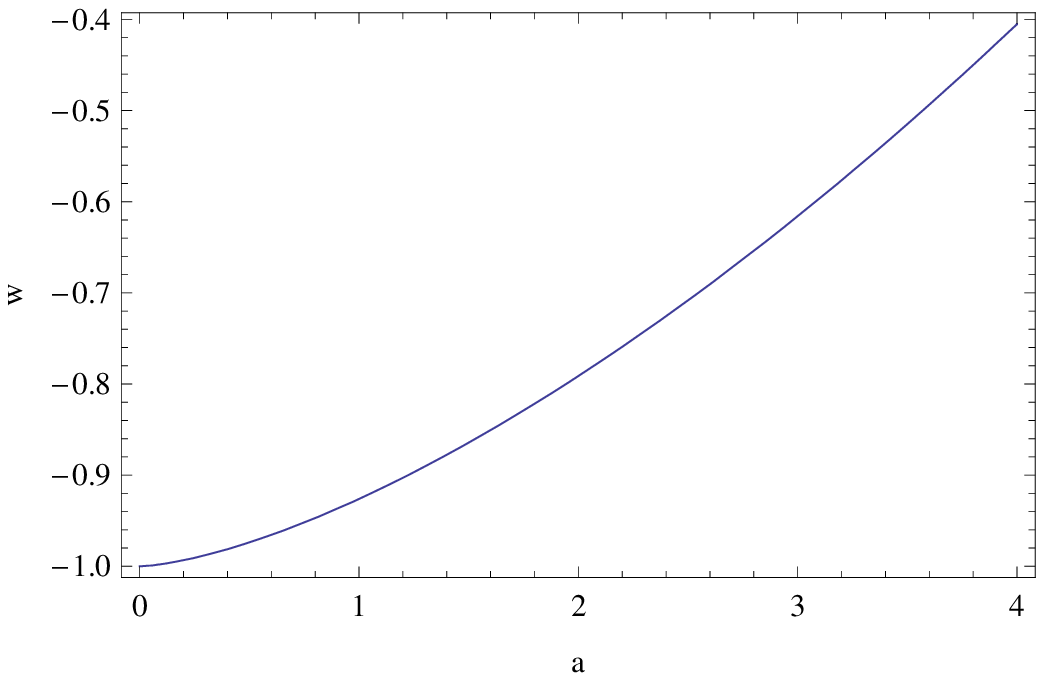}~\\
\vspace{1mm} ~~~~~~~~~~~~Fig.8~~~~~~~~~~~~~~~~~~~~~~~~~~~~~~~~~~~~~~~~~~~~~~~~~~~~~~~~Fig.9\\
\vspace{6mm} Figs. 8 and 9 show the evolution of $\eta$ and $w$
with increase in the scale factor $a$ when new agegraphic dark
energy
 is considered as a Ricci dark energy.

\vspace{6mm}

\end{figure}

Evolution of $\eta$ and $w$ are plotted against $a$ in figures 8
and 9 respectively. It is observed that $\eta$ is decreasing with
increase in the scale factor $a$. However, the equation of state
parameter of the new agegraphic dark energy is increasing with the
evolution of the universe when it is considered as Ricci dark
energy.

\section{\normalsize\bf{Conclusion }}
In this work we first studied the evolution of the equation of
state parameter for the Ricci dark energy in presence of dark
matter in non-interaction. We observed that depending on the value
of $c^{2}$ it evolves across the cosmological-constant boundary.
Subsequently, we studied the correspondence between Ricci dark
energy and tachyonic field. In a non-interacting situation, we
reconstructed the potential and scalar field for the tachyonic
field considering it as Ricci dark energy. Plotting the potential
against the scale factor $a$ and the scalar field $\phi$ we
observed that as the universe evolves, the potential of the
tachyonic field decreases when it is considered as Ricci dark
energy. Next we reconstructed the potential and scalar field of
the DBI-essence considering it as Ricci dark energy. In this case,
increase of the scalar field and decrease of the potential was
observed with the evolution of universe. Considering the new
agegraphic dark energy as Ricci dark energy we reconstructed the
equation of state parameter of new agegraphic dark energy and it
has been observed that the equation of state parameter of the new
agegraphic dark energy is increasing with the evolution of the
universe when it is considered as Ricci dark energy.
\\\\
{\bf References:}\\
\\
$[1]$ A.G. Riess, et al., {\it Astron. J.} {\bf 116} 1009 (1998);
S. Perlmutter, et al., {\it Astrophys. J.} {\bf 517} 565 (1999);
D.N. Spergel, et al.,  {\it Astrophys. J. Suppl.} {\bf 170}
377(2007); J.K.Adelman-McCarthy, et al., arXiv: 0707.3413.
\\
$[2]$ R. R. Caldwell, R. Dave, and P. J. Steinhardt, {\it Phys.
Rev. Lett.} {\bf 80} 1582 (1998).\\
$[3]$ S. Weinberg, in {\it Sources and Detection of Dark Matter
and Dark Energy in the Universe}, edited by D. B.
Cline (Springer, New York, 2001), pp. 18–26 (2000).\\
$[4]$ C. Gao, F. Wu, X. Chen and Y-G Shen, {\it Phys. Rev. D} {\bf
79} 043511 (2009)arXiv:0712.1394v4 [astro-ph]; X. Zhang, {\it
Phys. Rev. D} {\bf 79} 103509 (2009).\\
$[5]$ S. Chattopadhyay and U. Debnath, {\it Astrophys Space Sci}
{\bf 319} 183 (2009); J. Zhang, X. Zhang, and H. Liu, {\it Eur.
Phys. C} {\bf 52} 693 (2007); M. R. Setare, J. Zhang, and X.
Zhang, {\it J. Cosmol. Astropart. Phys.} 03 (2007) 007.\\
$[6]$ A. Sen, {\it JHEP} {\bf 065} 0207. \\
$[7]$ J. Martin and M. Yamaguchi, {\it Phys. Rev. D} {\bf D 77}
123508 (2008).\\
$[8]$ M. Setare , {\it Astrophys. Space Sci} DOI: 10.1007/s10509-009-0214-4 (2009).\\
$[9]$ C. Feng and X. Li, {\it Phys. Lett. B} {\bf 680} 355 (2009).\\
$[10]$ C.J. Feng, arXiv:0806.0673 [hep-th]; C.J. Feng, {\it Phys.
Lett. B} {\bf 670}231 (2008) ; C.J. Feng, {\it Phys. Lett. B} {\bf
672} 94 (2009) ; C.J. Feng, {\it Phys. Lett. B} {\bf 676} 168
(2009); C.J. Feng, X. Zhang, arXiv:0904.0045 [gr-qc]; C.J. Feng,
X.-Z. Li, arXiv:0904.2972 [hep-th];  L.N. Granda, A. Oliveros,
{\it Phys. Lett. B} {\bf 671} 199 (2009) .\\
$[11]$ M. R. Setare and J. Sadeghi, {\it Int J Theor Phys} {\bf 47} 3219 (2008).\\
$[12]$ T. Padmanabhan,  {\it Current Science} {\bf 88} 1057 (2005).\\
$[13]$ J. Martin and  M. Yamaguchi,  {\it Phys. Rev. D}
{\bf 77} 123508 (2008).\\
$[14]$ R. G. Cai, {\it Phys. Lett. B} {\bf 657} 228 (2007).\\
$[15]$ H. Wei and  R. G. Cai,  {\it Phys. Lett. B} {\bf
660} 113 (2008).\\

\end{document}